\documentclass[apjl]{emulateapj}

\usepackage{graphicx}
\usepackage{color}
\usepackage{amssymb}
\usepackage{verbatim}
\def\gr{$\gamma$-ray}

\shorttitle{Fermi/LAT observations of 1ES~0229+200: EGMF and EBL implications}

\begin{document}

\title{Fermi/LAT observations of 1ES~0229+200: implications for extragalactic magnetic fields and background light}

\author{Ie.~Vovk\altaffilmark{1}, A.~M.~Taylor\altaffilmark{1}, D.~Semikoz\altaffilmark{2,3} and A.~Neronov\altaffilmark{1}}

\altaffiltext{1}{ISDC Data Centre for Astrophysics, Ch. d'Ecogia 16, 1290, Versoix, Switzerland}
\altaffiltext{2}{APC, 10 rue Alice Domon et Leonie Duquet, F-75205 Paris Cedex 13, France}
\altaffiltext{3}{Institute for Nuclear Research RAS, 60th October Anniversary prosp. 7a, Moscow, 117312, Russia}

\begin{abstract}
We report the observation in the GeV band of the blazar 1ES~0229+200, which over recent years has become one of the primary sources used to put constraints on the Extragalactic Background Light (EBL) and Extragalactic Magnetic Field (EGMF). We derive constraints on both the EBL and EGMF from the combined Fermi-HESS data set taking into account the direct and cascade components of the source spectrum. We show that the limit on the EBL depends on the EGMF strength and vice versa. In particular, an EBL density twice as high as that derived by Franceschini et al. (2008) is allowed if the EGMF is strong enough. On the other hand, an EGMF strength as low as $6 \times 10^{-18}$~G is allowed if the EBL density is at the level of the lower bound from the direct source counts. We present the combined EBL and EGMF limits as an exclusion plot in two dimensional parameter space: EGMF strength vs. EBL density. 
\end{abstract}



\section{Introduction}

Very-high-energy (VHE) \gr\ flux from distant blazars is absorbed on the way from the source to the Earth through its interaction with the Extragalactic Background Light (EBL) photons (\cite{gould}). The measurement of the induced distortions of the VHE \gr\ flux from distant hard-spectrum blazars by the effect of absorption on the EBL was used to derive constraints on the EBL density (\cite{AhaEBL,Aha1ES0229,Orr_EBL_1ES0229}).

The conventional derivation of the upper bound on the EBL from \gr\ observations adopts the assumption that the intrinsic powerlaw-type spectrum of the primary source (a distant blazar) is characterized by the slope $dN_\gamma/dE\sim E^{-\Gamma}$ with $\Gamma \ge 1.5$. This assumption appears reasonable in the framework of the most simple synchrotron-self-Compton (SSC) models for the broad band spectra of blazars. However, particular blazars considered for the derivations of EBL limits (the blazars with hardest intrinsic \gr\ spectra) may well not fit into this simplest SSC model framework, so that the assumption of $\Gamma\ge 1.5$ might not be applicable (\cite{aharonian08,bottcher08,katarzinski06,Lefa_hard1,Lefa_hard2,neronov11}). If the intrinsic spectra of the blazars used for the derivation of the upper bound on the EBL density are harder than $\Gamma=1.5$, this upper limit is relaxed (\cite{mazin}).

Constraints on the intrinsic slope of the spectra of blazars  can be obtained from the observations by the Fermi Large Area Telescope (LAT) (\cite{atwood09}) in the energy band below $\sim 100$~GeV, where the effect of absorption on the EBL becomes negligible.  However, the blazars used for the derivation of constraints on the EBL are characterized by hard spectra, which makes it difficult to observe their flux below 100~GeV. In fact, the blazar 1ES~0229+200, which provides the tightest constraints on the EBL (\cite{Aha1ES0229}) is not  listed in the catalogue of sources detected by LAT in two-year exposure (\cite{fermi_catalog}), with only upper limits on the source flux derived from the LAT data (\cite{NeronovEGMF,Tavecchio:2010mk,TaylorEGMF}) and a weak detection reported by \cite{Orr_EBL_1ES0229}.

An additional difficulty for such constraints is that the spectrum of hard blazars might be composed of two contributions. Apart from the direct \gr\ emission from the primary source, an additional contribution is expected from the \gr\ cascade initiated in the intergalactic medium (IGM) by the absorbed VHE $\gamma$ rays (\cite{aharonian94,plaga95}). The overall flux and the spectral shape of the cascade contribution are determined by the strength of the extragalactic magnetic field (EGMF) (\cite{Aharonian_cascade,plaga95,neronov07,NerSem_prediction}). Uncertainty of the EGMF strength introduces an uncertainty of the importance of the cascade contribution and prevents the measurement of the slope of the intrinsic \gr\ spectrum of the source. In fact, the limits on the EBL derived up to now are based on an underlying assumption about the EGMF strength (the EGMF should be strong enough to suppress the cascade contribution up to the $\sim$TeV energy band), which is not justified a-priori. 

If the assumption  on the EGMF strength is relaxed, the \gr\ data can be used to measure the EGMF strength. The electron-positron pairs, created as a result of the absorption of multi-TeV photons, up-scatter the Cosmic microwave background (CMB) as they cool, creating secondary emission in the GeV domain (\cite{NerSem_prediction}). With the mean-free-path of TeV $\gamma$ rays being $D_\gamma \simeq 100~(E_\gamma/10~ {\rm TeV})^{-1}$~Mpc, the production of these multi-TeV energy $e^+$-$e^-$ pairs occurs predominantly in extragalactic space. Furthermore, if non-negligible magnetic fields are present in the region these pairs are born into, they deflect significantly from their initial directions, resulting in the secondary GeV photons having a reduced probability of reaching the observer. Thus, the apparent flux is suppressed at low energies. Observational evidence for the presence of such a suppression places constraints on the EGMF strength to be $\gtrsim 10^{-17}$~G (\cite{Dermer,Dolag:2010ni,NeronovEGMF,Tavecchio:2010mk,TaylorEGMF}).

This limit on the EGMF is derived assuming a certain level of EBL density (a low EBL density derived by \cite{Franceschini_EBL} is used in most of the publications on the EGMF in order to derive conservative bounds). However, a variety of EBL models do exist (\cite{gilmore09, kneiske04, stecker_ebl}), which spans the full range of the present uncertainty in the EBL density. This uncertainty therefore introduces an uncertainty into the bounds on the EGMF derived from the \gr\ data. Higher(lower) EBL density leads to stronger(weaker) absorption of the primary \gr\ flux along with a stronger(weaker) cascade contribution to the observed flux in the GeV range. The suppression of such stronger(weaker) cascade contribution would require a stronger(weaker) EGMF. 

Thus, the limits on the EBL derived from the \gr\ data depend on the assumptions made about the EGMF strength and vice versa. This implies that the correct procedure for the derivation of  limits on the EBL from the \gr\ data should include marginalization over the possible EGMF values. Conversely, the correct procedure for the derivation of the limits on the EGMF should include marginalization over the possible EBL densities and spectra. 

Practical implementation of this correct procedure for the derivation of the EBL and EGMF bounds, however, present a challenging task since the quality of the \gr\ data is usually insufficient for the exploration of the entire EBL-EGMF parameter space. Exploration of this parameter space requires the measurement of the source spectra both in the TeV and GeV energy ranges. 

In this Letter we report the observations of the blazar 1ES~0229+200 in the 1-300~GeV energy range using LAT's data with three years of exposure. The detection of the source below 100~GeV provides the information necessary for the correct analysis of the EGMF-dependent upper bound on the EBL density and of the EBL-dependent lower bound on the EGMF strength.  We present such bounds in the form of a two-dimensional exclusion plot in the ``EBL density'' vs. ``EGMF strength'' parameter space. 


\section{Analysis}

In this work we use publicly available data of the LAT instrument collected over the period from August 2008 till November 2011.  We use the Pass 7 data and analyse them using the Fermi Science Tools v9r23p1 software package, with the patches as of the November 1st 2011. We limit ourselves with the class 2 events, as is recommended by the Fermi/LAT team\footnote{http://fermi.gsfc.nasa.gov/ssc/data/analysis/documentation/\newline Cicerone/Cicerone\_Data\_Exploration/Data\_preparation.html}. We select only photons with energies in the 1-300~GeV range for the analysis. During the spectral fitting, we include in the analysis all the sources listed in the Fermi second year catalogue (\cite{fermi_catalog}) within $5^\circ$-circle around the position of 1ES~0229+200.

The fluxes in separate energy bins are computed with the spectral indices of all sources frozen at the values quoted in the Fermi/LAT second year catalogue; the spectral index of 1ES~0229+200 is fixed at the best-fit value in the 1-300~GeV band. For energies where 1ES~0229+200 is not detected in the bin, we compute the 90\%-confidence upper limit.

To compute the \textquotedblleft spectral butterfly\textquotedblright{} we scan the value of the likelihood in the \textquotedblleft index-normalization\textquotedblright{} parameter space and select the corresponding 68\% confidence region.

We employ a Monte-Carlo description of the electromagnetic cascades in order to determine the arriving spectra observed for the case of different strength of the EGMF (see \cite{TaylorEGMF} for more details). The shape of the  EBL spectrum is assumed to follow that derived by \cite{Franceschini_EBL}, while the normalization of the EBL is left free. A lower bound on the normalization of the EBL from the direct source counts (as summarized by \cite{dominguez11}) is $\simeq 15\%$ lower than the normalization of the EBL model of \cite{Franceschini_EBL}. Recent analysis of GeV to TeV spectra of several blazars (\cite{Orr_EBL_1ES0229}), however, suggested that the EBL level is somewhat higher, reaching $\simeq 60\%$ of the EBL model of \cite{Franceschini_EBL} at the 2$\sigma$ confidence level. Both these bounds are taken into account in our analysis.

In our calculations we consider the suppression of the cascade emission by time delay effects (\cite{plaga95}). The TeV \gr\ emission from the source is observed to be stable on the time scale of $\ge 3$~yr, from the initial HESS observations of the source (\cite{Aha1ES0229}) till the recent re-observation by the Veritas telescope (\cite{Dermer}). We do not consider the alternative possibility for the suppression of the cascade flux due to the extended nature of the cascade emission region. The generalization of our analysis to this case, however, is straightforward, the main difference being that the suppression of the cascade emission due to the source extension would result in a somewhat higher bound on the EGMF strength (\cite{TaylorEGMF}).

\begin{figure}
  \center \includegraphics[width=\linewidth]{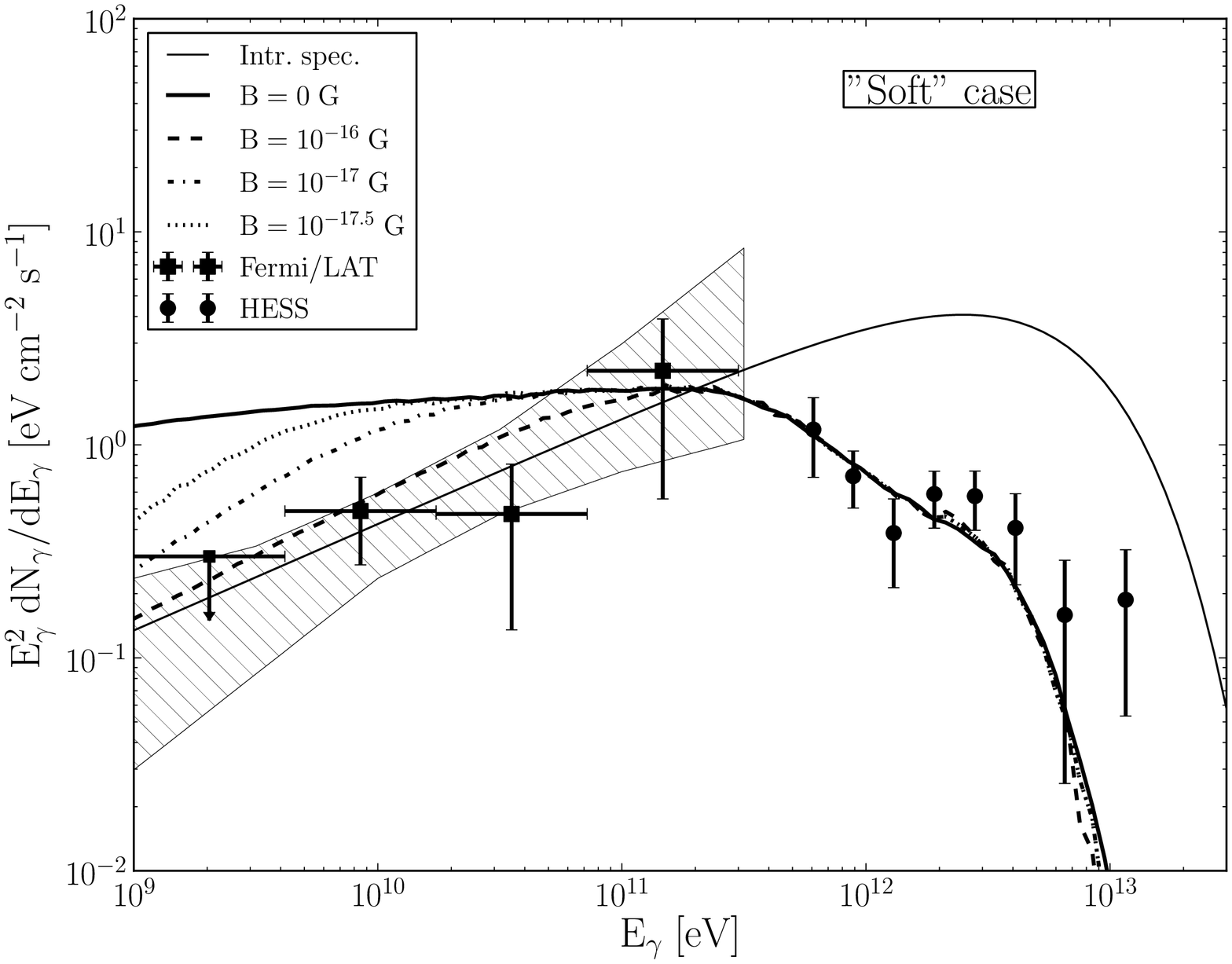}
  \center \includegraphics[width=\linewidth]{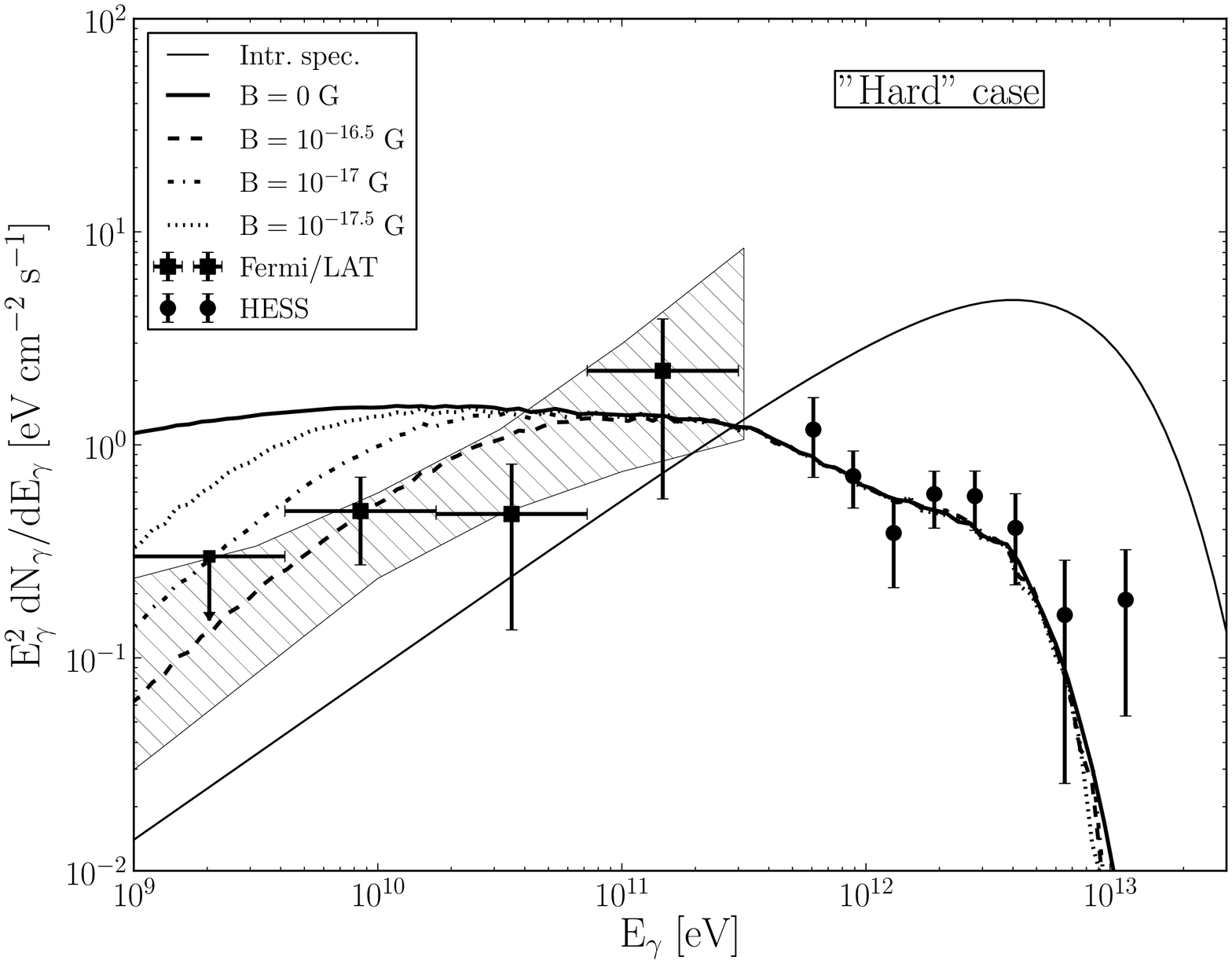}
  \caption{Top panel: the GeV-TeV SED under the assumption of a soft ($\Gamma=1.5$) intrinsic source spectrum. Bottom panel: the GeV-TeV SED under the assumption of a hard ($\Gamma=1.2$) intrinsic source spectrum. Lines for different values of EGMF strengths represent the sums of the intrinsic spectrum and the corresponding predicted cascade emission. In both cases the Fermi/LAT flux \textquotedblleft butterfly\textquotedblright { } shown corresponds to the 68\% confidence region.}
\label{soft}
\end{figure}

\section{Results}

1ES~0229+200 is detected by LAT in the 1-300~GeV energy band with significance $\simeq 7 \sigma$. The test statistics (TS) value found in the likelihood analysis is $TS=45$. Modelling the source spectrum as a powerlaw we find the slope of the spectrum $\Gamma=1.36\pm 0.25$ and normalization at 20~GeV  $(1.4 \pm 0.5) \times 10^{-15}$ (MeV$\cdot$cm${}^2 \cdot$sec)${}^{-1}$ (at the 68\% confidence level).  The spectrum of the source found from the LAT data is shown in Fig. \ref{soft} together with the HESS spectrum at higher energies. The source is not detected below $\simeq 3$~GeV, only an upper limit on the source flux can be derived in this energy band. 

As is discussed above, the source spectrum in the 3-300~GeV energy band can have two contributions: the direct \gr\ signal from the primary source and emission from the electromagnetic cascade developing in the IGM. It is not clear a-priori if the measured spectral slope, consistent with $\Gamma\simeq 1.5$, characterizes the intrinsic source spectrum, the spectrum of the cascade component, or comprises a summed spectrum of the two (similar in strength) contributions. For instance, a spectral index harder then 1.5, as found in the analysis of \cite{stecker_1ES0229} in the TeV band, would be
indicative of a GeV spectrum which results from the sum of both the intrinsic spectrum and that of the cascade. Different possibilities for the dominance of one of the two components in the spectrum are illustrated in the two panels of Fig.~\ref{soft}. In both models the normalization of the intrinsic spectrum is chosen to fit the HESS measurements in the TeV band. We also assume that the intrinsic source spectrum has a high-energy cut-off at $E_{cut}=5$~TeV. As it was shown by \cite{TaylorEGMF}, this choice minimizes the strength of the cascade contribution in the Fermi/LAT energy band. 

In the upper panel the main contribution to the 3-300~GeV source flux is given by the direct flux of the primary source, shown by the thin solid line. This is possible only if  the cascade component is suppressed by the influence of a strong enough EGMF. If the EGMF is negligible, the flux of the direct and cascade emission (thick solid line) will largely dominate over the direct emission.  Strong EGMF ($\ge 10^{-17}$~G) is needed to sufficiently suppress the cascade emission down to the level of the error bars of the LAT measurements in the 3-300~GeV range.

If the EGMF is weaker than $\sim 3\times 10^{-17}$~G, the cascade emission provides the dominant contribution to the source spectrum, as is illustrated in the lower panel of Fig. \ref{soft}. The only possibility to make the LAT measurement consistent with observations is to assume that the intrinsic spectrum of the primary source has a slope harder than $\Gamma=1.5$. The hardness of the intrinsic source spectrum depends on the EGMF strength. For the particular example shown in the lower panel of Fig. \ref{soft}, the assumption that the EGMF strength $B\le 3\times 10^{-17}$~G imposes a constraint on the intrinsic source spectrum $\Gamma\le 1.2$. In fact, if the intrinsic source spectrum is even harder, the intrinsic source flux contribution to the 3-300~GeV band flux becomes negligible and the flux is completely dominated by the cascade emission.

\begin{figure}
  \center \includegraphics[width=\linewidth]{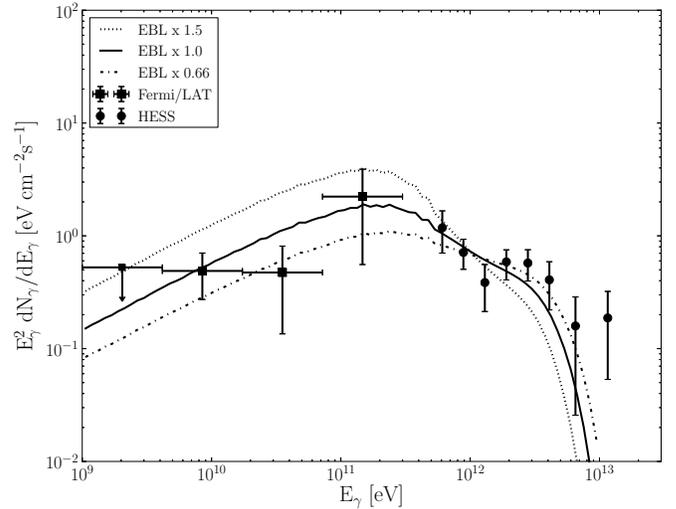}
  \caption{GeV-TeV SED under the assumption of a soft ($\Gamma=1.5$) intrinsic spectrum for a different EBL scales. The EGMF strength of $10^{-16}$~G is adopted here, and the normalization of the spectra is chosen to fit the HESS data. The reference EBL scale is that of \cite{Franceschini_EBL}.}
\label{EBL_scaling}
\end{figure}

The overall normalization of the cascade emission is determined by the density of the EBL. An increase of the EBL density leads to the stronger absorption of multi-TeV $\gamma$ rays and, consequently, to stronger cascade emission. 
To the contrary, reducing the EBL normalization down to the level of the lower bound from the direct source counts opens up the possibility of a weaker EGMF, down to $\sim 6 \times 10^{-18}$~G. The effect of changing the EBL normalization is illustrated in Fig. \ref{EBL_scaling}. In this figure a spectral slope of $\Gamma=1.5$ and EGMF of $10^{-16}$~G have been adopted, and three different (66\%, 100\% and 150\% in terms of the EBL level reported by \cite{Franceschini_EBL}) levels of EBL have been used.

The maximal normalization of the EBL which can still be consistent with the data depends on the strength of the EGMF. Too strong an EBL can result in a large over-prediction of the strength of the cascade emission, even after taking into account the suppression of this emission by the EGMF effects. Thus, the upper bound on the EBL derivable from the \gr\ observations of 1ES~0229+200 is EGMF-dependent. 

\begin{figure}
 \centering \includegraphics[width=\linewidth]{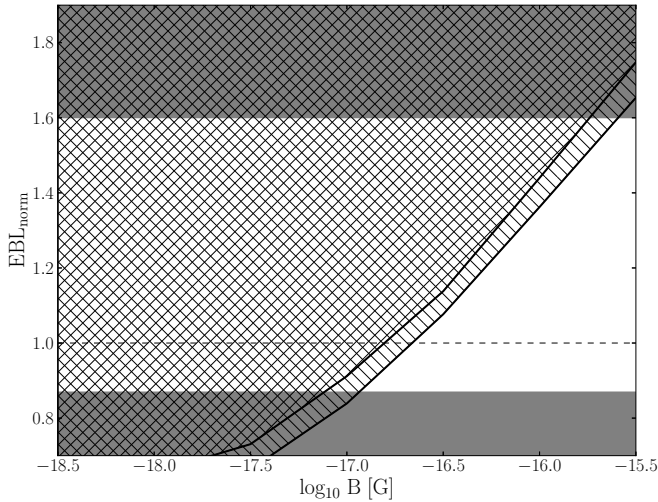}
 \caption{Allowed normalizations of the \cite{Franceschini_EBL} EBL spectrum, as a function of the EGMF strength. Dark grey shaded regions are restrictions from \cite{dominguez11} and \cite{Orr_EBL_1ES0229}. The single stroke (cross) hatched region is excluded by the analysis presented here at the 95\% (99.7\%) confidence level.}
 \label{EBL_exclusion}
\end{figure}

In order to find this bound, we compute the allowed ranges of the EBL normalization for a set of EGMF strengths and intrinsic spectrum spectral indices. The range of the EGMF strengths scanned over lay in the range $3 \times 10^{-19}$~G to $10^{-14}$~G, while the spectral indices were varied within the range from 1.5 to 0. We then find the best-fit set of values in this \textquotedblleft EGMF-EBL-$\Gamma$\textquotedblright parameter space, and chose an appropriate confidence region. The projection of this region onto the \textquotedblleft EGMF-EBL\textquotedblright plane is shown in Fig.~\ref{EBL_exclusion} and comprises the not hatched part of the plot. The hatched part thus represents the EBL-dependent bound on the EGMF (or, equivalently, the EGMF-dependent bound on the EBL).

One can see from this figure that for EGMF strengths $B\sim 10^{-17}$~G the upper bound sits at the level derived by \cite{Franceschini_EBL}. If the EGMF is at the level of $\sim 10^{-15}$~G, the allowed EBL normalization is by a factor of 2 higher than that of the \cite{Franceschini_EBL} and \cite{dominguez11} models. 

\section{Conclusion}
Following our investigations into the newly detected blazar 1ES~0229+200 with Fermi/LAT we find that, if the \cite{Franceschini_EBL} EBL level is adopted, an EGMF strength of at least $10^{-17}$~G is required in order for the inevitable GeV cascade spectral component to be consistent with observations. However, we find that, more generally, the EGMF lower bound is dependent on the EBL level adopted, as demonstrated in Fig.~\ref{EBL_exclusion}.

One should keep in mind that this bound is derived for large ($\lambda_B > 1$~Mpc) correlation lengths of the EGMF, and scales approximately as $\lambda_B^{-1/2}$ for $\lambda_B \lesssim 1$~Mpc. Our result lies in agreement with previous findings (\cite{TaylorEGMF,Huan:2011kp}). Under the assumption that the source is stable on a time-scale much longer than 3 years, a larger bound on the EGMF is obtained (\cite{NeronovEGMF,TaylorEGMF,Dolag:2010ni,Tavecchio:2010mk}). 

We also find that an intrinsic blazar spectral index of $\Gamma=1.5$ is able to sit in agreement with observations of 1ES~0229+200 by Fermi/LAT ($\Gamma=1.36\pm 0.25$). Thus our results can find agreement with the underlying assumptions behind previous EBL constraint calculations by \cite{AhaEBL}. However, the remaining uncertainty in the origin of the blazar's spectral shape in the GeV domain leaves open the possibility for a harder intrinsic spectrum ($\Gamma<1.5$).

We note, that in previous works of \cite{AhaEBL} and \cite{Orr_EBL_1ES0229} the authors made the implicit assumption for the absence of the cascade contribution in the GeV-TeV domain, which is equivalent to the presence of the strong EGMF with $B \gtrsim 10^{-15}$~G. As can be seen from Fig. \ref{EBL_exclusion}, for such a strong field our result is also compatible with the level of EBL found by \cite{Orr_EBL_1ES0229}. For weaker fields, $B \sim 10^{-17}$~G, we are also compatible with the findings of \cite{dominguez11}. However, for such weak fields a proper account should be given of the GeV-TeV cascade component contribution.

Future observations with ground-base Cherenkov telescopes, such as MAGIC Stereo, HESS 2 and CTA, will dramatically improve the measurements of blazar spectra below $\sim 100$~GeV, allowing much better constraints on the EGMF and EBL. In the particular case of 1ES~0229+200, if the intrinsic spectrum is hard, the presented EGMF lower bound may be transformed to a \textit{measurement}, helping to better understand the nature of the extragalactic magnetic field.


\acknowledgments
This work is supported in part by the SCOPES project No.~IZ73Z0\_128040 of the Swiss National Science Foundation.


\end{document}